\documentclass[useAMS,usenatbib]{mn2e}
\usepackage{times}
\usepackage{graphicx,subfigure,epsf}
\usepackage{amssymb}

\def \ref {\noindent\hangindent=1.0in\hangafter=1}

\def\ltsima{$\; \buildrel < \over \sim \;$}
\def\simlt{\lower.5ex\hbox{\ltsima}} 
\def\gtsima{$\; \buildrel > \over \sim \;$}
\def\simgt{\lower.5ex\hbox{\gtsima}} 

\title[Hubble Space Telescope 
ultraviolet spectroscopy of blazars: emission lines properties and black hole masses]{Hubble Space Telescope 
ultraviolet spectroscopy of blazars: emission lines properties and black hole masses}
\author[E. Pian, R. Falomo, A. Treves]{E. Pian$^{1}$\thanks{E-mail:
pian@ts.astro.it (EP); falomo@pd.astro.it (RF); Aldo.Treves@uninsubria.it (AT)}, 
R. Falomo$^{2}$, A. Treves$^{3}$\\
$^{1}$INAF, Astronomical Observatory of Trieste, Via G.B. Tiepolo 11, I-34131
Trieste, Italy\\
$^{2}$INAF, Astronomical Observatory of Padova, via dell'Osservatorio 5, I-35122
Padova, Italy\\
$^{3}$Department of Physics and Mathematics, University of Insubria, Via Valleggio 11, I-22100 Como,
Italy}

\begin{document}

\date{Accepted ... Received 2005 January 20; in original form 2005 January 1}

\pagerange{\pageref{firstpage}--\pageref{lastpage}} \pubyear{2005}

\maketitle

\label{firstpage}

\begin{abstract}
The ultraviolet (UV)  spectra of 16 blazars ($<z>  \simeq 1$) from the
archives of the Hubble  Space Telescope Faint Object Spectrograph have
been analyzed in order to study  in a systematic way the properties of
their broad  UV emission lines. We  find that the  luminosities of the
most prominent  and intense lines,  Ly$\alpha$ and C~IV~$\lambda$1549,
are  similar  to those  of  normal  radio-loud  quasars at  comparable
redshifts.   However,  the  equivalent  widths  of  blazar  lines  are
significantly  smaller than those  of radio-loud  quasars.  Therefore,
while the intrinsic broad line region luminosity of blazars appears to
be indistinguishable from that  of radio-loud quasars, their continuum
must  be  comparatively  higher,  most probably  due  to  relativistic
beaming.  We  have  combined  the  UV luminosities  of  the  de-beamed
continuum with  the emitting gas  velocity to derive estimates  of the
masses of the central supermassive black holes.  The size of the broad
line region was computed in two ways: 1) via an empirical relationship
between UV  continuum luminosity  and broad line  region size,  and 2)
through  the external  photon  density required  by  blazar models  to
reproduce the inverse Compton  components observed at gamma-rays.  The
second method  yields significantly  different results from  the first
method, suggesting  that it provides only  a very rough  estimate or a
lower limit  on the size of the  broad line region.  We  find that the
average mass of  the central black holes in blazars  is $\sim 2.8 \times
10^8 M_\odot$,  with a large dispersion, comparable  to those computed
for other radio-loud active galactic nuclei.
\end{abstract}


\section{Introduction}

Highly Polarized Quasars (HPQ, also referred to as Flat Spectrum Radio
Quasars) and, occasionally, BL Lacertae objects, collectively known as
blazars, exhibit  broad emission  lines superimposed on  their optical
and ultraviolet (UV) continua (Netzer et al.  1994; Scarpa, Falomo, \&
Pian 1995;  Vermeulen et  al.  1995; Corbett  et al.  1996;  Scarpa \&
Falomo 1997; Koratkar et al. 1998; Pian et al. 2002; D'Elia, Padovani,
\&  Landt  2003).   Emission  lines  play an  important  role  in  the
energetics of blazars: some  models of multiwavelength blazar emission
(Dermer \& Schlickeiser 1993; Sikora et al.  1994; Ghisellini \& Madau
1996) predict  that the broad  line region  (BLR) photons  are Compton
upscattered to X- and gamma-ray energies by the relativistic particles
composing  the jet  plasma,  and form  luminous  high energy  spectral
components, which often dominate  the overall blazar output (Mattox et
al.   1997;  Bloom et  al.   1997; Wehrle  et  al.   1998; Hartman  et
al. 2001; Ballo et al. 2002; Pian et al. 2002). The role of broad line
emission in  shaping the spectrum  of different classes of  blazars is
however not  fully assessed (Fossati  et al.  1998; Ghisellini  et al.
1998; Padovani et al. 2003).

More in general, the characteristics of the BLR of blazars may help in
investigating  the  interplay  between  the  accretion  disk  and  the
relativistic jet,  which is more  prominent in blazars than  in normal
QSO and  Seyferts (Celotti  et al. 1997;  Maraschi \&  Tavecchio 2003;
D'Elia et al. 2003; Wang, Luo,  \& Ho 2004).  AGN broad emission lines
may  be  used also  to  estimate the  masses  of  the compact  objects
residing in the nuclear  centers, most likely supermassive black holes
(BH), by  exploiting their dynamical  effect on the  line-emitting gas
clouds.  The application of the  virial theorem requires that the size
of  the BLR is  determined either  with direct  methods (reverberation
mapping  technique,  
Peterson \&  Wandel  2000;  Kaspi  et al.  2000;
Peterson et al. 2004), or  with indirect arguments (Kaspi et al. 2000;
Vestergaard 2002). In the  latter case the uncertainties are obviously
larger,  and the  methods must  be tested  carefully.  Since  this has
important consequences  on the evolution and demographics  of AGNs, it
is  crucial to  accomplish these  measurements both  for low  and high
redshift sources.  This  approach has been adopted in  the estimate of
BH masses of large samples  of quasars based on optical emission lines
(Woo \& Urry 2002; McLure \& Dunlop 2004).  The advantage of using UV,
rather than optical emission lines  is that the former correspond to a
higher   ionization   state   and   are  therefore   presumably   more
representative of the dynamics close to the central massive object.

In this  paper we  present the  analysis of the  broad and  intense UV
emission  lines  of   16  blazars observed  by  the  Hubble  Space
Telescope (HST)  and the Faint Object Spectrograph  (FOS).  Previous studies of
AGN UV spectra have been  carried out  by Bechtold  et al.   (2002), who
focussed on  the absorption systems,  by Kuraszkiewicz et  al.  (2002)
and Kuraszkiewicz et al. (2004),  who present a complete HST FOS atlas
of emission line parameters of AGNs, and by Evans and Koratkar (2004),
who recalibrated the pre-COSTAR AGN spectra. We concentrate here on the UV spectra
of blazars. Based on  the radiative and  kinematic properties  of the
broad emission line region (BLR), we measure the luminosities of their
BLRs  and derive  estimates of  the BLR  sizes and  of the  central BH
masses.

We   adopt   the   ``concordant   cosmology'',   $\Omega_m$   =   0.3,
$\Omega_\Lambda$ = 0.7,  and assume $H_0$ = 72  km s$^{-1}$ Mpc$^{-1}$
(Spergel et al. 2003).  Luminosities reported by other authors and used 
in this paper have been transformed into this cosmology.

\section{Sample selection and data analysis}

We  have  retrieved  from  the HST  archive\footnote{using  MAST,  the
Multi-mission  Archive at  STScI, see  http://archive.stsci.edu.}  all
pre- and post-COSTAR  FOS grating spectra of sources  previously 
classified as blazars (Wall \& Peacock 1985; Impey  \& Tapia 1988;
Impey \& Tapia 1990; Impey et al. 1991; Stickel et al. 1991; Stocke et
al. 1991; Padovani \& Urry 1992; 
Wills et al. 1992;  Perlman et al. 1996).   We also included in
our   final  list   PKS~1229-021,   which,  despite   having  low
polarization (Wills et al. 1992), is considered a blazar because of 
significant emission at MeV-GeV frequencies 
(Hartman  et  al.  1999),  and   3C~273,  which  has
intermediate  properties  between  those  of blazars  (strong  radio
emission,  superluminal motion,  gamma-ray emission, jet emission dominance
at hard X-ray energies, e.g., Haardt et al. 1998; Grandi \& Palumbo 2004),  and those  of 
Seyfert galaxies (broad emission lines, big blue bump).  
We selected spectra taken  with high-resolution gratings in  the UV region
(G130H, G190H,  G270H, G400H).   

This search yielded 24 objects with measurable spectra.  Spectra taken
with  the  same grating  within  one  day  were averaged  to
increase the signal-to-noise ratio.   For this work we have considered
only the 16 sources  with significant (larger than 3$\sigma$) emission
line detections. These are reported in Table~1. 

For 6  sources there are also low-resolution grating (G160L) observations 
in the archive, obtained nearly  simultaneously to the high-resolution 
spectra (i.e., within one day).  Since the line
parameters and spectral indices derived from the former are not significantly 
different from those measured in the high-resolution spectra, we have neglected
these spectra.

Although the considered objects  do  not  represent a
complete sample,  they form a sizeable dataset to investigate
the UV line properties of blazars.


\begin{table*}
\begin{minipage}{140mm}
\caption{Parameters of Blazars observed with HST FOS}\label{foslog}
\begin{tabular}{@{}cccccccc@{}}
\hline
Object      & Alt. Name & $z$  & $E_{B-V}^a$ & Date & Range$^b$ & $\alpha_\lambda^c$ & $f_{1350}^d$ \\
\hline
BL0403-1316 &          & 0.571 & 0.058  & 11.4 Oct 1991 & 1570-4780 & $1.20 \pm 0.04$ & 1.98 \\
BL0420-0127 &          & 0.915 & 0.125  & 23.5 Dec 1996 & 2220-3280 & $0.00 \pm 0.22$ & 0.80 \\
BL0537-4406 &          & 0.896 & 0.037  & 16.7 Sep 1993 & 2220-3280 & $0.07 \pm 0.20$ & 1.64 \\
BL0637-7513 &          & 0.656 & 0.095  & 25.6 May 1992 & 1610-3270 & $1.25 \pm 0.06$ & 5.64 \\
BL0954+5537 & 4C 55.17 & 0.901 & 0.0088 & 20.9 Jan 1993 & 1570-3280 & $0.31 \pm 0.08$ & 0.60 \\
BL1144-3755 &          & 1.048 & 0.097  & 15.6 Jul 1993 & 2220-3280 & $0.20 \pm 0.23$ & 1.42 \\
BL1156+2931 &          & 0.729 & 0.019  & 26.7 Feb 1995 & 1620-4780 & $0.52 \pm 0.04$ & 8.29 \\
BL1226+0219 & 3C 273   & 0.158 & 0.02   & 16.8 Jan 1991 & 1090-3280 & $1.68 \pm 0.04$ & 245  \\
            &          &       &        & 09.4 Jul 1991 & 1290-3290 & $1.35 \pm 0.06$ & 117  \\
BL1229-0207 &          & 1.045 & 0.032  & 01.1 Jan 1995 & 1570-3280 & $0.81 \pm 0.12$ & 2.58 \\
BL1253-0531 & 3C 279   & 0.538 & 0.028  & 08.5 Apr 1992 & 1570-4780 & $0.18 \pm 0.03$ & 1.54 \\
BL1611+3420 & DA 406   & 1.401 & 0.018  & 04.9 Apr 1992 & 2220-4780 & $0.91 \pm 0.08$ & 0.81 \\
BL1641+3954 & 3C 345   & 0.595 & 0.013  & 07.9 Jun 1992 & 1600-4770 & $0.65 \pm 0.04$ & 3.03 \\
            &          &       &        & 20.4 Aug 1995 & 1570-4800 & $0.96 \pm 0.16$ & 0.43 \\
BL2223-0512 & 3C 446   & 1.404 & 0.075  & 11.8 Sep 1991 & 2220-4780 & $0.00 \pm 0.11$ & 0.35 \\
BL2230+1128 & CTA 102  & 1.037 & 0.072  & 12.1 Sep 1991 & 2220-4780 & $1.13 \pm 0.06$ & 1.95 \\
BL2243-1222 &          & 0.63  & 0.051  & 09.4 Oct 1993 & 1610-3270 & $1.59 \pm 0.07$ & 5.18 \\
BL2251+1552 & 3C 454.3 & 0.859 & 0.105  & 11.9 Sep 1991 & 1620-4770 & $0.90 \pm 0.04$ & 2.86 \\
            &          &       &        & 15.3 Nov 1991 & 1620-3270 & $0.51 \pm 0.10$ & 4.35 \\
            &          &       &        & 19.4 Aug 1995 & 1700-4800 & $1.08 \pm 0.09$ & 1.72 \\
\hline
\multicolumn{8}{l}{$^a$ From the maps of Schlegel et al. (1998).} \\
\multicolumn{8}{l}{$^b$ Observed wavelength range, in \AA.} \\ 
\multicolumn{8}{l}{$^c$ Spectral index of the power-law 
fitted to the dereddened spectra ($f_\lambda \propto \lambda^{-\alpha}$).} \\
\multicolumn{8}{l}{$^d$ Normalization of the dereddened power-law continuum at 1350 \AA\ (rest frame),
in $10^{-15}$ erg~s$^{-1}$~cm$^{-2}$~\AA$^{-1}$.} \\
\end{tabular}
\end{minipage}
\end{table*}



\begin{table*}
\begin{minipage}{140mm}
\caption{Emission Line Measurements}\label{foslog}
\begin{tabular}{@{}cccccccc@{}}
\hline
\hline
&&&&&&& \\
&&&&&&& \\
&&&&&&& \\
\multicolumn{8}{l}{$^a$ {\bf This table is provided on page 9, in landscape 
format}} \\
&&&&&&& \\
&&&&&&& \\
&&&&&&& \\
\hline
\end{tabular}
\end{minipage}
\end{table*}



\begin{table*}
\centering
\begin{minipage}{140mm}
\caption{Composite ultraviolet blazar spectrum.}\label{compparam}
\begin{tabular}{@{}cccccccc@{}}
\hline
Line & EW$^a$ & FWHM$^a$  & Rel. Intensity$^b$ & Ratio$^c$ & Ratio (LBQS)$^d$ & Ratio (RLQa)$^e$ & Ratio (RLQb)$^f$ \\
\hline
Ly$\beta$  & $6.5 \pm 0.5$ & 10  & $8.9 \pm 0.5$  & 10.6 & 9.3  & 19   & 19.1  \\              
Ly$\alpha$ & $71 \pm 3$    & 22  & $84 \pm 3$     & 100  & 100  & 100  & 100   \\
SiIV       & $5.2 \pm 0.7$ & 21  & $5.6 \pm 0.7$  & 6.6  & 19   & 6.8  & 8.6   \\
C IV       & $45 \pm 2$    & 20  & $45 \pm 1$     & 53   & 63   & 66   & 52    \\
C III]     & $7 \pm 1$     & 23  & $6.6 \pm 0.7$  & 7.8  & 29   & 11   & 13.2  \\ 
Mg II      & $19 \pm 2$    & 30  & $16 \pm 1$     & 19   & 34   & 24   & 22.3  \\ 
\hline
\multicolumn{8}{l}{$^a$ At rest frame, in \AA.  Uncertainties are about $\sim$15\%.} \\ 
\multicolumn{8}{l}{$^b$ Obtained from EW and continuum normalized to the flux at 1500 \AA\ (rest frame).} \\ 
\multicolumn{8}{l}{$^c$ Average percentage intensity ratio with respect to Ly$\alpha$ as resulting from our measurements.}\\
\multicolumn{8}{l}{$^d$ Same as in Col. 5 for 718 objects in the Large Bright Quasar Survey (Francis et al. 1991).}\\
\multicolumn{8}{l}{$^e$ Same as in Col. 5 for for 60 radio loud quasars (Zheng et al. 1997).}\\
\multicolumn{8}{l}{$^f$ Same as in Col. 5 for 107 radio loud quasars (Telfer et al. 2002).}\\
\end{tabular}
\end{minipage}
\end{table*}


After applying a correction for the Galactic absorption using the maps
of Schlegel, Finkbeiner  and Davis (1998) and the  extinction curve of
Cardelli, Clayton and Mathis (1989), we measured the equivalent widths
(EWs),  the intensities  and the  full  width at  half maximum  (FWHM)
values of the emission lines  fitting a linear local continuum on each
side of the  line (see Table~2).  The EW  uncertainties were estimated
by assuming 2$\sigma$ variations of the local continuum.

For   each   object   we   have   combined the   spectra   taken
quasi-simultaneously (within 1  day) with different gratings, excluded
the regions  affected by emission  or absorption features,  binned the
signal  in 20-50 \AA\  wavelength intervals  and fitted  the continuum
with  a power-law.  To account  for calibration  uncertainties  of the
data, we added a 5\%  systematic error to the statistical errors.  The
derived power-law  spectral indices and flux  normalizations are given
in Table~1.

A  similar analysis  of the  continuum  and line  properties of  these
objects  has  been  presented  in  Kuraszkiewicz  et  al.  (2002)  and
Kuraszkiewicz et al. (2004).

\section{Results}

The redshifts  of our  objects (see Table~1)  range between $z = 0.158$ and
$z = 1.404$,  with an  average value of  $<z> =  0.84 \pm  0.31$. Thus,  the lines
typically  detected  in our  FOS  spectra  are Ly$\beta$,  Ly$\alpha$,
C~IV~$\lambda$1549,  C~III]~$\lambda$1909, Si~IV~$\lambda$1400  and in
some  cases Mg~II~$\lambda$2798.   We report  in Table~1  the spectral
indices and  normalizations of the UV  continua and in  Table~2 the EW
and the FWHM values of the emission lines.

In 3  cases (3C~273, 3C~345,  and 3C~454.3) observations at  more than
one  epoch  are available. Variations of line  and  continuum emission 
are observed, with maximum amplitudes of factors of $\sim$2 and 7, respectively.  
However the observations are too limited and sparse to allow 
a meaningful assessment of correlated line and continuum variability.

In  section 3.1  we describe  the average  properties of  the emission
lines of blazars in the UV  spectral region; in section 3.2 we use the
line and continuum properties of the UV spectra to estimate the masses
of the central BHs.

\subsection{Emission line properties}

In  order to produce  a representative  high signal-to-noise  ratio UV
spectrum  of  blazars,  we  have  combined  all  the UV spectra  in  our
dataset. Each spectrum was first reduced to  rest-frame and then
normalized  to its  average continuum  flux.  The  resulting composite
blazar spectrum,  normalized to unity  at the reference  wavelength of
1500 \AA, is shown in  Fig.~1.  The EW, relative intensities, and FWHM
of the emission lines of the composite spectrum are given in Table~3.

\begin{figure*}
\centering
\vspace{-4cm}
\includegraphics[scale=0.8,angle=0,keepaspectratio]{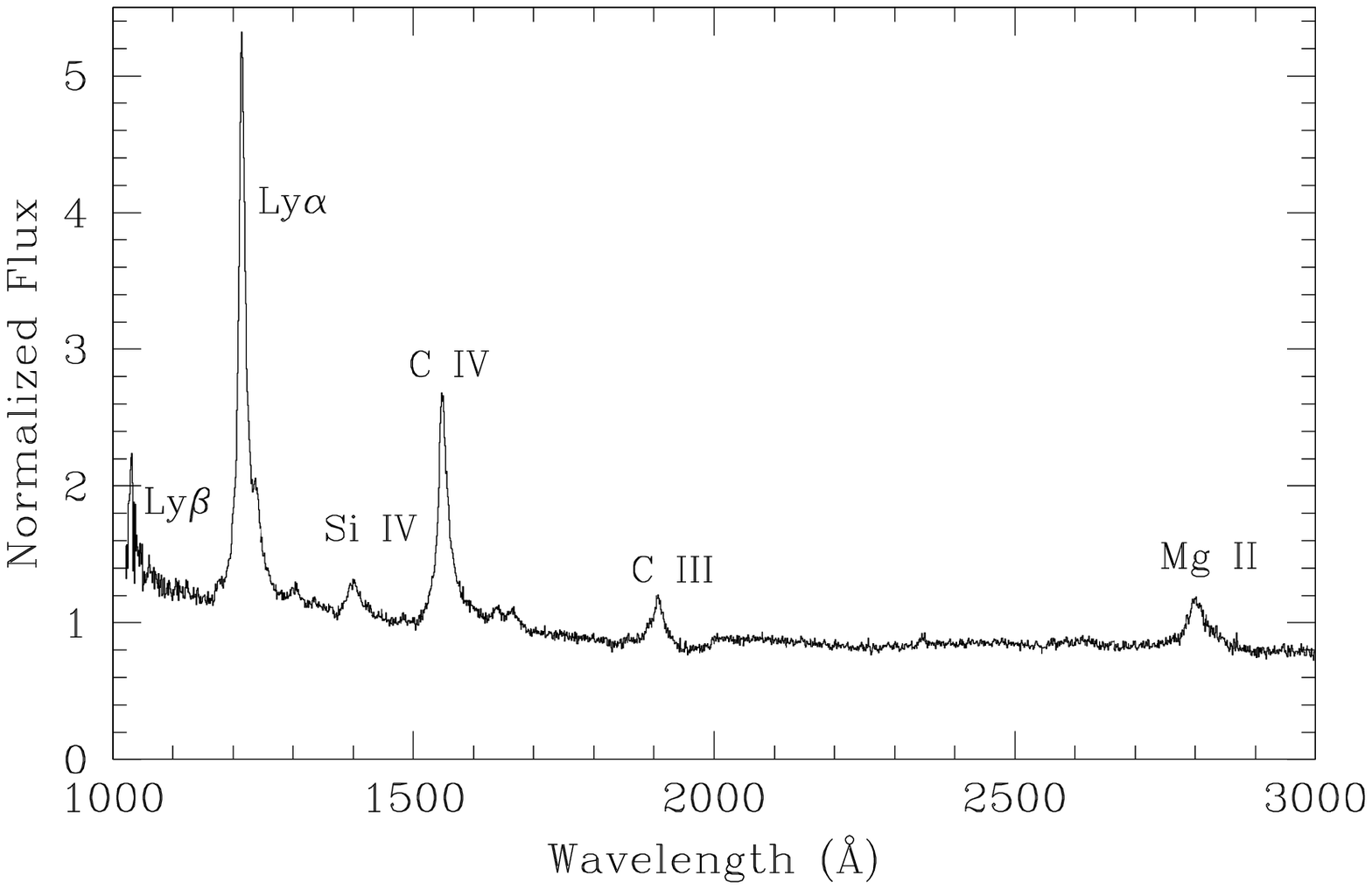}
 \caption{Composite UV spectrum of blazars obtained from the average of 16 spectra. 
The most prominent emission lines are labeled.}
\label{compuvblsp} 
\end{figure*}

The composite  spectrum of blazars is  similar to that  of normal QSO.
The line ratios of blazars  and normal AGNs are also not significantly
different.   This  is  illustrated  in  Fig.~2, where  we  report  the
luminosities  of  the   Ly$\alpha$  and  C~IV~$\lambda$1549  lines  of
blazars, compared  with those  of a list  of radio-loud  quasars (RLQ)
observed by  HST FOS (Wills et  al.  1995). Note that in this list of RLQ
there are 8  objects  in common  with  our sample of blazars; therefore,
for the purpose of the comparison, these 8 sources have been considered as
blazars and have been excluded from the RLQ list.
We also compare with the RLQ 3C~390
(1845+79,  $z =  0.056$), which  has hybrid  properties, i.e.,  it has
substantial  polarization   (1.3\%,  Impey   et  al.  1991),   but  is
lobe-dominated at radio wavelengths (Ghisellini et al.  1993).  Figure
2   shows   that   the   intensity   ratio  of   Ly$\alpha$   vs   the
C~IV~$\lambda$1549  in blazars  is consistent  with that  exhibited by
normal RLQ.  

\begin{figure}
\centering
\includegraphics[scale=0.47,angle=0,keepaspectratio]{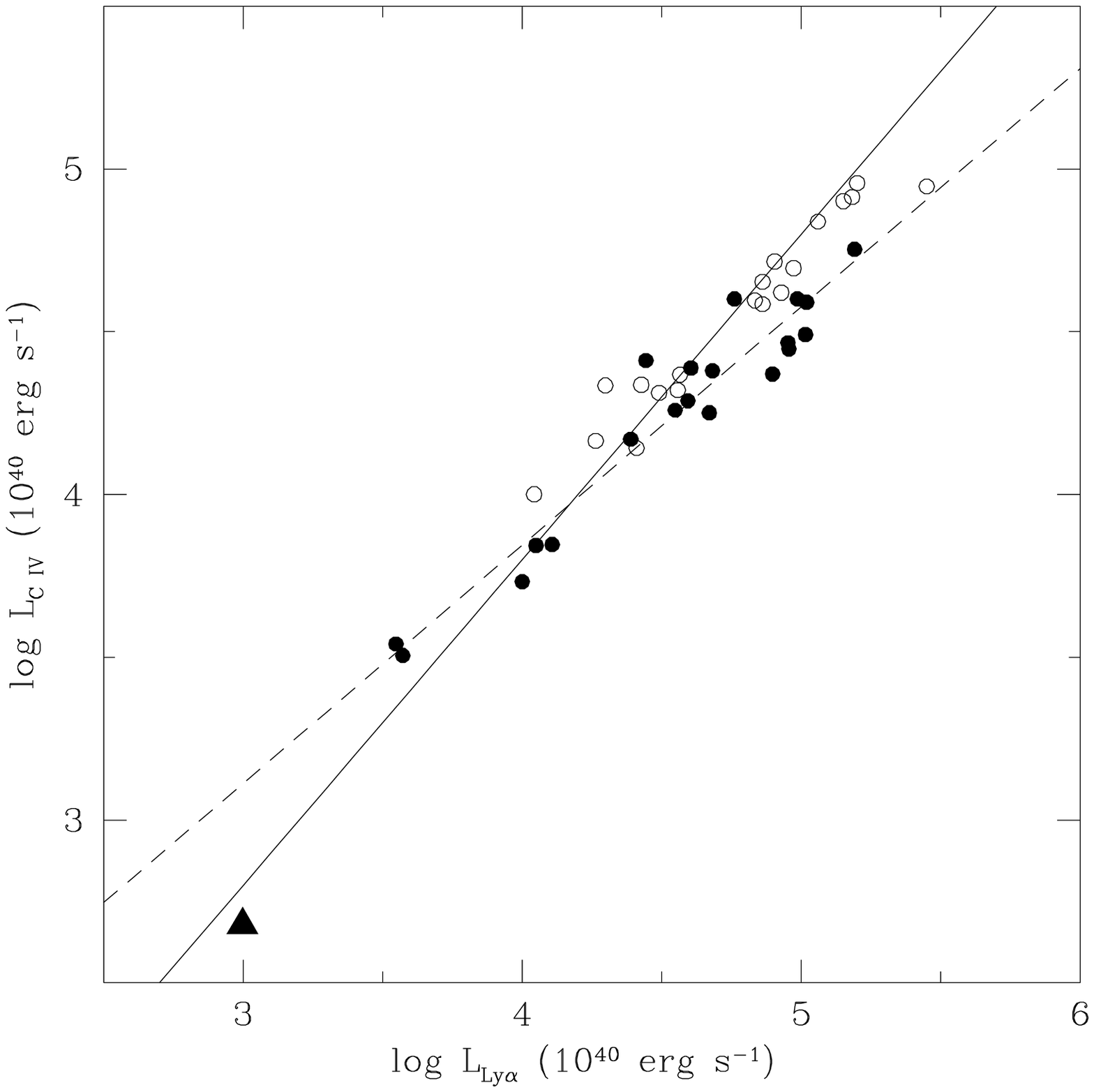}
\caption{Luminosities of   C~IV~$\lambda$1549   and  Ly$\alpha$
emission  lines for  the blazars  in our  list (filled circles). 
Line measurements of  a given source at multiple  epochs have not been
averaged.   The   Ly$\alpha$  and  C~IV~$\lambda$1549   emission  line
intensities   have  been  normalized   to  their   respective  average
values.  The   solid  line  is  the   expected  C~IV~$\lambda$1549  to
Ly$\alpha$    intensity    ratio    (L(C~IV~$\lambda$1549) = 
0.63$\times$L(Ly$\alpha$),  e.g.  Francis  et al.  1991). The  dashed line  is the
least  square regression  line L(C~IV~$\lambda$1549) = 
8.3$\times$L(Ly$\alpha$)$^{0.73}$). For comparison, the line luminosities 
of 19 RLQ (open circles) from Wills et al. (1995) and of 3C~390 (triangle)
are also shown.}
\label{lineratio} 
\end{figure}

More in general,  we have compared in Table~3 the average
intensity  ratios of  the  lines  we detect  in  our composite  blazar
spectrum with  those reported by  other authors for larger  samples of
QSO or  RLQ (Francis et al. 1991; Zheng et al. 1997; Telfer et al. 2002).  
Except for the C~III]~$\lambda$1909 line,  which appears
somewhat  underluminous   in  blazars,  there   is  good  overall
agreement.  This  comparison suggests that the  structure and physical
state of the BLR in blazars and normal RLQ are indistinguishable.  This is also
confirmed by the comparison of the Ly$\alpha$ line luminosities.  
In  Fig.~3 we report  the continuum  luminosity at 1350 \AA\ as a function  of the
Ly$\alpha$ luminosity for blazars  and RLQ.  With the exception of 3C~390, which is at
relatively low redshift, the blazars and RLQ 
have a similar range  of Ly$\alpha$ luminosities. The averages 
of the logarithmic
distributions, in erg~s$^{-1}$, are $<{\rm log L(Ly\alpha)}> = 44.55
\pm 0.11$  and $<{\rm log  L(Ly\alpha)}> = 44.72  \pm 0.12$ for  blazars and
RLQ, respectively (the uncertainties represent the errors associated with the averages, i.e. the
standard deviations divided by the square root of the number of objects). 
However, due to relativistic beaming, blazars have
more luminous continua than RLQ, i.e., blazar lines have smaller EW (see Fig.~3).
While RLQ emission lines, for any continuum luminosity, have EW between 100 \AA\ 
and 1000 \AA, part of the blazars exhibit line EW between 10 \AA\ and 100 \AA, and these
have the most strongly boosted continua. From comparison with the RLQ, we have estimated 
the luminosity enhancement due to beaming.

Relativistic aberration affects the non-thermal synchrotron luminosity
and  depends  on  the  fourth  power  (for  a  jet  geometry)  of  the
relativistic  Doppler  factor $\delta$  (=  [$\Gamma  (1  - \beta  cos
\theta)]^{-1}$).
RLQ are thought to be the  parent population of blazars: their jets 
are directed away from the line of sight, so that their 
luminosities are only weakly affected by relativistic beaming.  
Therefore, we have
estimated  the beaming amplification for  the blazars  by  assuming that
their continuum luminosities should exhibit a dependence on Ly$\alpha$
line luminosities similar to that  of RLQ (Fig.~3).  We fitted the RLQ
line (in erg s$^{-1}$) and continuum (in erg s$^{-1}$ \AA$^{-1}$) luminosities 
to a power-law and obtained the dependence: 

\begin{equation}
L(1350 {\rm \AA}) = 0.46 \times 10^{-3}  L(Ly\alpha)^{1.02},
\label{kuv.eq}
\end{equation}

which has a  scatter of 0.2dex in L(1350 \AA).   We will use this relationship 
in Section 3.2.1 to 
correct the continuum luminosities of blazars for the beaming.

\begin{figure}
\centering
\includegraphics[scale=0.47,angle=0,keepaspectratio]{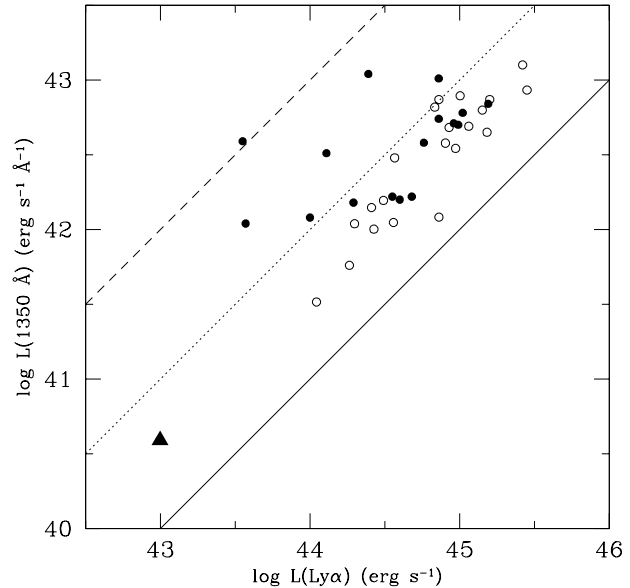}
\caption{Ly$\alpha$ luminosity vs continuum luminosity at 1350 \AA\ (rest frame) 
for blazars (filled
circles) and RLQs (open circles). The triangle represents the low redshift RLQ 3C~390.
The lines are the loci of the EW = 10 \AA\ (dashed),
100 \AA\ (dotted) and 1000 \AA\  (solid).}
\label{blvsrlq} 
\end{figure}

\subsection{Black hole masses}

Under the  assumption that the dominant mechanism  responsible for the
width of  the broad emission  lines is the gravitational  potential of
the  central supermassive  BH, and  that the  line widths  reflect the
Keplerian  velocities of  the line-emitting  material in  a virialized
system  (Wandel  et al.  1999;  McLure \&  Dunlop  2001), the BH  mass
M$_{BH}$ is given by:

\begin{equation}
M_{BH} = G^{-1} v^2 R_{\rm BLR} 
\label{mld.eq}
\end{equation}

where  $v$ is the  velocity of  the gas  gravitationally bound  to the
central  BH, $R_{\rm  BLR}$  is the  size of  the  BLR, and  G is  the
gravitational  constant.  The  velocity $v$  can be  obtained directly
from the FWHM  of the broad emission lines ($v  = f \times v_{FWHM}$),
where $f$ is  a factor that depends on the  geometry and kinematics of
the BLR (e.g., McLure \& Dunlop 2002; Vestergaard 2002).

\subsubsection{Size of the BLR}

The  most  reliable  method   to  derive  $R_{\rm  BLR}$ is  through
the reverberation  mapping technique  (e.g.   Peterson et  al. 2004,  and
references  therein).  This uses the  time lag  of the  emission line
light curve with respect to the continuum light curve to determine the
light crossing size of the BLR in AGNs.  However, this method requires
intensive monitoring of  the UV continuum and of the  lines and can be
used  only for  a  limited number  of  objects
(e.g., Korista et al. 1995; Onken et al. 2002), including  one of  our
sources, 3C~273 Paltani \& T\"urler 2005). An
alternative  way to  estimate $R_{\rm  BLR}$ is  to use  the empirical
relationship  found between  $R_{\rm BLR}$  and the  optical continuum
luminosity (Kaspi et al.  2000).

We have  derived this  relationship in the  UV for  a sample of  15 PG
quasars  and 10  Seyfert 1  galaxies having  BLR radii  determined via
reverberation  mapping in the optical (Kaspi  et al.  2000) and  measured  UV spectra
(Vestergaard  2002).   One of the PG quasars is 3C~273, which is also a 
member of our blazar sample.  For this object, we corrected the continuum luminosity
for the beaming effect (see Eq.~1).  For  one  of  the Seyferts,  NGC~4151,  the  UV
continuum  of which  varies with  high amplitude,  we  re-measured the
continuum  luminosity at  1350  \AA\ from  the  average IUE  spectrum.
These quantities are reported in  Fig.~4.  We fitted the BLR radii and
the luminosities at  1350  \AA\ (rest frame) with a  power-law  and obtained  the
following relationship:

\begin{equation}
R_{BLR} = (22.4 \pm 0.8) \left(\frac{\lambda L_\lambda(1350 {\rm \AA})}{\rm 10^{44} erg~s^{-1}}\right)^{0.61 \pm 0.02} {\rm lt-days} 
\label{kuv.eq}
\end{equation}

\noindent
If we exclude 3C~273 from the fit, the result is unchanged (the power-law index is 
$0.60 \pm 0.02$ in this case).
We note that the slope in Eq.~3 is consistent with that determined in the same wavelength range
by Kaspi et al. (2005, index $0.56 \pm 0.05$), it is
slightly flatter than that found in the optical
by Kaspi et al. (2000, index $0.70 \pm 0.03$) and slightly steeper than that 
found by  McLure and Jarvis  (2002, index $0.50 \pm 0.02$) at  3000 \AA,  based on
a very similar sample  of PG
quasars and Seyfert galaxies. 

\begin{figure}
\centering
\includegraphics[scale=0.47,angle=0,keepaspectratio]{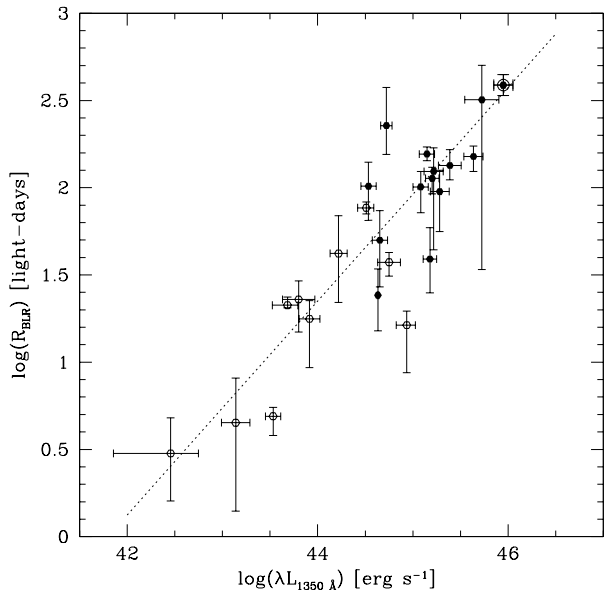}
\caption{Size of the BLR as a function of the continuum power output at
1350 \AA\ (rest frame) for the sample of PG quasars (filled circles) and 
Seyfert 1 galaxies (open circles) of Kaspi et al. (2000). The encircled point on the top right
represents 3C~273. The luminosities have been taken from Vestergaard (2002). 
The line fitting the 2 quantities (Eq.~3) is also
reported (dotted).}
\label{blvsrlq} 
\end{figure}

Since the relationship between the BLR radius and continuum luminosity
is supposed to  be valid in the case of a  thermal continuum (see also
discussion in Paltani  and T\"urler 2005), we must  correct the blazar
UV  continuum luminosities  for  the effect  of relativistic  beaming.

For the blazars with
continuum  luminosity   exceeding  the power-law dependence between the RLQ line and
continuum luminosities (Eq.~1),   we  adopted  the
continuum luminosities  computed with Eq.~1 at  the corresponding line
luminosity,  and  derived the  BLR  radii  through  Eq.~3.  These  are
reported  in Table~4. The  correction of  the continuum  luminosity is
relevant (i.e., larger  than $\sim$3 times the scatter) for 4 objects  (see
Fig.~3). We  note that  our estimate  of the BLR  radius of  3C~273 is
consistent with that reported by Paltani  and T\"urler (2005).

An  alternative,  independent  method  for  evaluating  $R_{\rm  BLR}$
consists in  coupling the luminosity  of the BLR with  the information
carried by the multiwavelength spectrum of the blazar.  Since blazars,
among  all  AGNs, are  the  only ones  with  a  spectrum extending  to
gamma-rays, this method is specific for the blazar class of AGNs.

Ten  of our  blazars have  multiwavelength energy  distributions which
have  been  fitted  with  synchrotron and  inverse  Compton  radiation
components (Ghisellini  et al. 1998).  The  latter component dominates
at the X- and gamma-ray energies and originates from the scattering of
relativistic electrons  off both synchrotron photons  (internal to the
jet) and external radiation fields.  These include broad line photons,
the  density  of  which,  $U_{ext}$,  is thus  estimated  through  the
multiwavelength spectral fit.

Following the  procedure adopted  by Celotti, Padovani  and Ghisellini
(1997), we reconstructed  the total luminosity of the  BLR for each of
our sources by using the intensities of the observed UV emission lines
and by assuming for the unobserved lines the line ratios of an average
quasar spectrum (Francis et al. 1991).  These derived BLR luminosities
are reported in Table~4.  

From the {\it fitted} densities  of the external photons $U_{ext}$ and
from  the  {\it observed}  BLR  luminosities,  the  size of  the  BLR,
$R_{BLR}$, can be derived according to:

\begin{equation}
R_{\rm BLR} = \sqrt{\frac{L_{\rm BLR}} {4 \pi c U_{\rm ext} \delta^2}}
\end{equation}

where  $\delta$ is  the relativistic  Doppler factor  required  by the
multiwavelength  modeling.   The BLR  radii  computed  with Eq.~4  are
reported  in  Table~4.   We  have  identified this  second  method  as
``spectral energy distribution (SED) method'', in order to distinguish
it from the one based on the empirical determination of $R_{BLR}$ from
the continuum  luminosity.  No clear correlation is  found between the
BLR radii  determined with the two methods.   One probable explanation
of  the  discrepancy is  that  the radiation  density  of  the BLR  is
generally   smaller  than  the   parameter $U_{ext}$    obtained  with
multiwavelength fits. This parameter includes not only  the
BLR photons, but also additional contributions, 
such as photons  coming directly from the accretion disk,
or  produced by  the dusty  torus, or  by larger  regions of  the jet
(Ghisellini priv.  comm.).  Thus,  the SED method may underestimate the
BLR sizes (and therefore the  BH masses) in some cases.  Moreover, the
uncertainties  associated  with  the  SED fit  parameters  are  large.
Therefore, although  we had  proposed the ``SED''  method for  BH mass
determination  in   an  individual  source   (PKS~0537--441,  Pian  et
al. 2002), it appears that this method cannot be generalized.

\subsubsection{Mass estimates}

In order to evaluate the BH  masses of our objects we have used Eqs.~2
and 3, adopting a standard value of $f = \sqrt{3}/2$ for the kinematic
factor, corresponding  to an isotropic distribution of  the BLR clouds
(Wandel 1999; Kaspi et al. 2000; Vestergaard 2002), that yields virial
BH masses consistent with  those derived from the M$_{BH}$-L$_{bulge}$
relationship  (Labita, Falomo,  \& Treves,  in prep.).   After setting
$v_{FWHM}$ to suitable units we obtain the relation:

\begin{equation}
M_{BH} = 3.26 \times 10^6
\left(\frac{\lambda L_\lambda(1350 {\rm \AA})}{\rm 10^{44} erg~s^{-1}}\right)^{0.61}
\left(\frac{v_{\rm FWHM}}{10^3~ \rm km~s^{-1}}\right)^2 M_\odot
\label{mv2.eq}
\end{equation}

Vestergaard (2002)  also derives  a formula for  the central  BH mass,
based on  the continuum measurements  at the rest frame  wavelength of
1350~\AA\  of  a  sample  of  26  AGN with  BLR  radii  determined  by
reverberation  mapping.  However,  this  relationship  was  calibrated
against the  BH masses determined  from optical measurements,  and not
directly from the BLR size, as we do.

By using Eq.~5 with the beaming-corrected L(1350 \AA) luminosities and
$v_{\rm  FWHM}$  estimated from  our  spectra,  we  have computed  the
central BH  masses.  For consistency  with Vestergaard (2002)  we have
used  the FWHM of  C~IV~$\lambda$1549.  These  masses are  reported in
Column 5 of Table~4, and in Fig.~5, where they are compared with those
computed  with  our  Eq.~5 for  a  sample  of  PG quasars,  for  which
Vestergaard (2002) reports UV luminosities and C~IV~$\lambda$1549 line
FWHM values.  The blazar BH  masses are statistically  consistent with
those of quasars: the averages of the logarithmic distributions,  
in solar masses, are $<{\rm log M_{BH}}> = 8.31 \pm 0.10$  and 
$<{\rm log M_{BH}}> = 8.42 \pm 0.08$ for  blazars and
RLQ, respectively (the uncertainties represent the errors associated 
with the averages, i.e. the
standard deviations divided by the square root of the number of objects). 

For  comparison, we  have also  computed  the blazar  BH masses  using
Vestergaard's  relationship  (Eq.~8  of  Vestergaard 2002),  with  our
measured blazar luminosities and FWHM values of the C~IV~$\lambda$1549
line.   These are  systematically  lower than  the corresponding  ones
determined using Eq.~5, by a  factor 1.4-2.  We have also compared our
estimated  BH masses  with those  determined for  the same  blazars by
other authors: a number of our  objects are in common with the samples
of Liang and Liu (2003), Woo  and Urry (2002), and Wang et al. (2004).   
Their  mass estimates have been reported in the last
3 columns of Table~4: our masses are generally smaller. This may 
be partially due to our correction of the continuum luminosity for the
beaming effect.  In particular, this must be the case for two of the four 
objects 
in our sample with the most strongly beamed continuum, 3C~279 and 0537-441.

We do  not find  a correlation of  the BH  mass with redshift  for our
blazar sample.

\begin{figure}
\centering
\includegraphics[scale=0.47,angle=0,keepaspectratio]{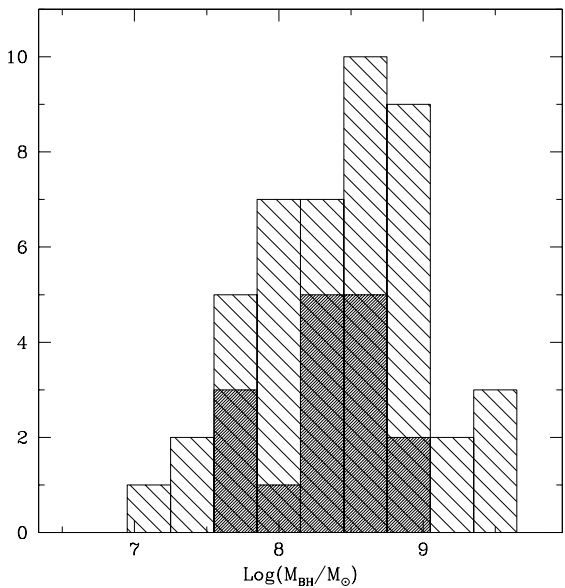}
\caption{Histograms of central BH  masses of AGNs.  The simply hatched
area represents the masses of  PG quasars computed with Eq.~5 from the luminosities
and C~IV~$\lambda$1549 emission line
FWHM values reported by from Vestergaard (2002).  The
double-hatched  area  represents the BH masses  of  our  blazar  sample.
Mass estimates obtained for a same source at different epochs have been averaged.}
\label{mbhhist} 
\end{figure}


\begin{table*}
\centering
\begin{minipage}{140mm}
\caption{BLR Luminosities and Central Black Hole Masses.}\label{rblrmbh}
\begin{tabular}{@{}cccccccc@{}}
\hline
Object & $L_{BLR}^a$ & $R_{BLR}^b$ & $R_{BLR,SED}^c$ & $M_{BH}^d$ & $M_{BH,LL}^e$ & $M_{BH,WU}^f$ & $M_{BH,W}^g$ \\
\hline                                      
 BL0403-1316 & 22.6 & 145 & ... & 2.4 &      & 12  &    \\
 BL0420-0127 & 18.7 & 145 & 438 & 2.3 & 8    & 11  & 9  \\
 BL0537-4406 & 6.93 &  77 & 211 & 0.5 & 16   &     & 5  \\
 BL0637-7513 & 50.2 & 285 & ... & 1.9 &      & 26  &    \\
 BL0954+5537 & 5.38 &  66 & 312 & 0.5 & 8    & 1.2 &    \\
 BL1144-3755 & 2.45 &  34 & ... & 0.4 &      &     &    \\
 BL1156+2931 & 13.7 & 115 & 158 & 4.3 & 8    &     &    \\
 BL1226+0219 & 33.8 & 219 & 17  & 4.0 & 0.2  &     & 16 \\
 BL1229-0207 & 74.1 & 357 & 117 & 7.4 & 10   &     &    \\
 BL1253-0531 & 2.42 &  36 & 104 & 0.8 & 4    & 2.7 & 3  \\
 BL1611+3420 & 34.1 & 196 & 92  & 6.2 & 40   & 37  &    \\
 BL1641+3954 & 12.5 &  92 & ... & 2.7 &      & 26  & 28 \\
 BL2223-0512 & 25.2 & 150 & ... & 2.8 &      &     &  6 \\
 BL2230+1128 & 41.4 & 260 & 33  & 3.1 & 12.6 &     &    \\
 BL2243-1222 & 47.8 & 271 & ... & 2.1 &      &     &    \\ 
 BL2251+1552 & 33.3 & 223 & 29  & 3.1 & 25   & 15  & 13 \\    
\hline
\multicolumn{8}{l}{$^a$ Luminosity of the BLR, in units of $10^{44}$ erg~s$^{-1}$.} \\
\multicolumn{8}{l}{$^b$ Size of the BLR computed via Eq.~3, in light days.} \\
\multicolumn{8}{l}{$^c$ Size of the BLR computed using the SED method, (Eq.~4), in light days.}  \\ 
\multicolumn{8}{l}{$^d$ Mass of the central BH computed using Eq.~5, in units of $10^8 M_\odot$. The statistical uncertainties, dominated}  \\ 
\multicolumn{8}{l}{ ~~~ by the scatter of Eq.~3, are about a factor 2.} \\
\multicolumn{8}{l}{$^e$ Mass of the central BH from Liang \& Liu (2003).}  \\ 
\multicolumn{8}{l}{$^f$ Mass of the central BH from Woo \& Urry (2002).}  \\ 
\multicolumn{8}{l}{$^g$ Mass of the central BH from Wang et al. (2004).}  \\ 
\end{tabular}
\end{minipage}
\end{table*}


\section{Summary and Conclusions}

We have  studied the properties of  the UV emission  lines of blazars,
mostly from  single epoch HST FOS  spectra, and found  that the average
blazar UV spectrum  is similar to that  of RLQ.  This is the  sum of a
thermal  and non-thermal component.   Our targets  are mainly  HPQ and
Low-Frequency Peaked BL Lacs (Padovani  \& Giommi 1995; Fossati et al.
1998) where  the  emission of  the  non-thermal synchrotron  component
peaks   at  optical/IR   frequencies. Therefore   a   large  relative
contribution from  the thermal  accretion disk is  expected in  the UV
region (e.g., Bregman et al.  1986).  This is clearly evident from the
spectral  energy   distribution  of  3C~273  (Ulrich   et  al.   1980;
Courvoisier  1998) and  may  be  significant in  3C~279  (Pian et  al.
1999).

With  the aim  of estimating  central BH  masses of  blazars,  we have
assumed Keplerian conditions in the  BLR gas motion and have evaluated
the BLR  size using the  results of a  fit of UV luminosities  and BLR
radii  of   a  sample  of   QSOs  having  BLR  sizes   determined  via
reverberation mapping in the optical.  We have derived a relationship between 
$R_{BLR}$ and luminosity in the UV domain (1350 \AA\ at rest frame), 
which exhibits a slope consistent with that of Kaspi et al. (2005), although
slightly flatter and steeper than proposed in the optical
and near-UV by Kaspi et al. (2000) and McLure \& Jarvis (2002), respectively.

For those 10 blazars having multiwavelength spectral fits we have also
applied  an   independent  method  of  BLR   size  determination, based on
the
combination of  the observed  $L_{BLR}$ and fitted  external radiation
density  (``SED'' method).   We  have not  found  a clear  correlation
between the BLR sizes obtained  with the two methods, with the largest
deviations observed  in the  sense of a  deficit of the  ``SED'' radii
with respect to  those obtained with the empirical  $R_{BLR} - \lambda
L_\lambda$  relationship.  We conclude  that  the  SED  method yields
a BLR size inconsistent with that derived from the continuum luminosity.

Our  estimated  BH masses  have  an average  of  $(2.8 \pm 2.0) \times  10^8
M_\odot$ (the quoted uncertainty is the standard deviation) 
and are comparable  with those of  lower redshift blazars,  estimated with
different  methods (Barth,  Ho,  Sargent 2003;  Falomo, Carangelo,  \&
Treves 2003).

The  distribution of  our  blazar  BH masses computed with Eq.~5 
is  consistent with  the
distribution  of the PG  quasar masses computed with the same equation.
These
results suggest  that the  differences between radio  powerful sources
and radio-weak  ones are not  due to the  mass of the BHs  residing at
their  centers.   However, the  validity  of  this  conclusion at  the
intermediate/high redshifts  must be  corroborated by the  analysis of
wider  samples of homogeneous  datasets.  Moreover,  further intensive
spectroscopic monitoring  of the brightest  blazars at optical  and UV
wavelengths is required, in  order to construct well sampled continuum
and emission  line light curves  for the application  of the reverberation
mapping technique.


\section*{Acknowledgments}

We  thank  R.  Bohlin  for  assistance  with  HST  data  analysis,
G. Ghisellini  for helpful discussion, and the referee, A. Koratkar, for a 
constructive report.  We acknowledge the use  of the
SIMBAD and  NED databases, publicly  available online.  This  work was
partially  supported by the  Italian Space  Agency under  the contract
I/R/056/02.



\begin{figure*}
\centering
\includegraphics[scale=0.8,angle=0,keepaspectratio]{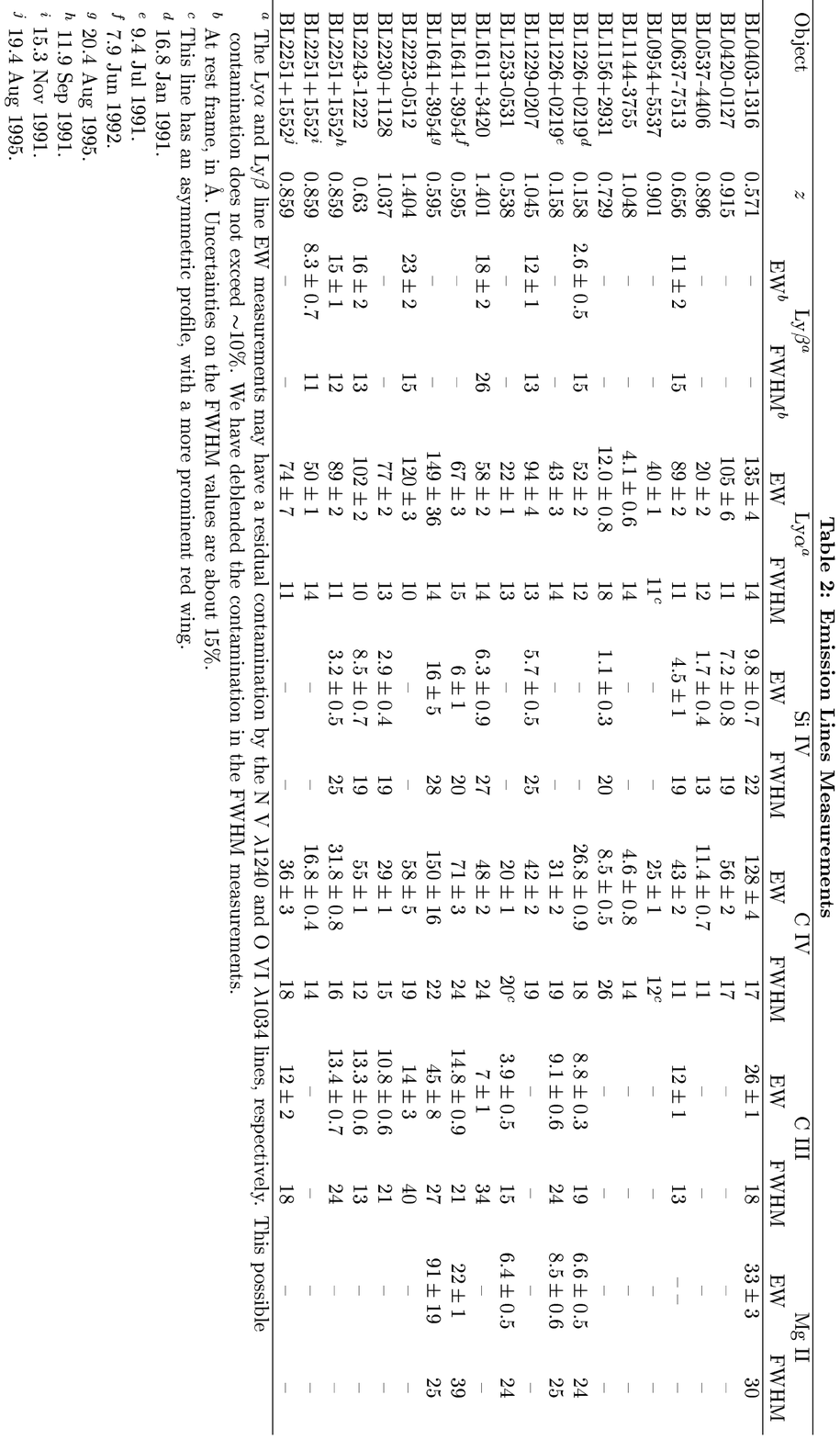}
 \caption{}
\end{figure*}

\bsp

\label{lastpage}

\end{document}